\pdfoutput=1
\documentclass[final, 5p, twocolumn, longtitle]{elsarticle}
\usepackage{geometry} %
\usepackage[version=4]{mhchem}
\usepackage{xcolor}
\usepackage{url}
\usepackage{adjustbox}
\usepackage[separate-uncertainty,retain-explicit-plus,per-mode = symbol, detect-weight=true, range-phrase=-, range-units=single]{siunitx}
\DeclareSIUnit{\keV}{\kilo\electronvolt}
\DeclareSIUnit{\MeV}{\mega\electronvolt}
\newcommand{\ncount}[1]{\num[group-minimum-digits=4]{#1}}
\usepackage[switch]{lineno}
\usepackage[hidelinks]{hyperref}
\usepackage{orcidlink}

\newcommand{\isotope}[2]{\ce{^{#2}#1}}
\newcommand{\xe}{\isotope{Xe}{136}}
\newcommand{\bi}{\isotope{Bi}{214}}
\newcommand{\po}{\isotope{Po}{214}}
\newcommand{\rn}{\isotope{Rn}{222}}
\newcommand{\qbb}{Q$_{\beta\beta}$}
\newcommand{\nldbd}{$0\nu\beta\beta$}

\date{} %

\begin{document}
\makeatletter
\long\def\MaketitleBox{%
  \resetTitleCounters
  \def\baselinestretch{1}%
  \begin{\elsarticletitlealign}%
   \def\baselinestretch{1}%
    \Large\@title\par\vskip18pt
    \normalsize\elsauthors\par\vskip10pt
    \footnotesize\itshape\elsaddress\par
    \penalty-10000
    \hrule\vskip12pt
    \ifvoid\absbox\else\unvbox\absbox\par\vskip10pt\fi
    \ifvoid\keybox\else\unvbox\keybox\par\vskip10pt\fi
    \hrule\vskip12pt
    \end{\elsarticletitlealign}%
}
\makeatother

\journal{Nuclear Instruments and Methods A}
\begin{frontmatter}
\title{Simulation and Rejection of Radon Progeny Backgrounds in nEXO}

\author[Yale]{M.~P.~Watts\orcidlink{0000-0002-9708-0056}}
\author[PNNL]{B.~T.~Foust\orcidlink{0000-0003-1713-3128}}
\author[Yale]{G.~Richardson\orcidlink{0000-0001-9353-2791}\fnref{fn1}}
\author[Drexel]{E.~T.~Chambers\fnref{fn2}}
\author[PNNL]{R.~Saldanha\orcidlink{0000-0003-2771-3281}}
\author[Yale]{D.~C.~Moore\orcidlink{0000-0002-2358-4761}}
\author[SUBATECH,UMass]{A.~Amy\orcidlink{0009-0007-9075-4277}}
\author[SLAC]{A.~Anker}
\author[PNNL]{I.~J.~Arnquist\orcidlink{0000-0002-5643-8330}}
\author[Drexel]{A.~Atencio}
% ...

\address[Yale]{Wright Laboratory, Department of Physics, Yale University, New Haven, CT 06511, USA}
\address[PNNL]{Pacific Northwest National Laboratory, Richland, WA 99352, USA}
\address[Drexel]{Department of Physics, Drexel University, Philadelphia, PA 19104, USA}
\address[SUBATECH]{SUBATECH, Nantes Universit\'e, IMT Atlantique, CNRS/IN2P3, Nantes 44307, France}
\address[UMass]{Amherst Center for Fundamental Interactions and Physics Department, University of Massachusetts, Amherst, MA 01003, USA}
\address[SLAC]{SLAC National Accelerator Laboratory, Menlo Park, CA 94025, USA}
% ...

\fntext[fn1]{Now at Facility for Rare Isotope Beams, Michigan State University, East Lansing, MI 48824, USA}
\fntext[fn2]{Now at University of Hawaii at Manoa, Honolulu, HI 96822, USA}
\fntext[fn3]{Now at Research Center for Neutrino Science, Tohoku University, Sendai 980-8578, Japan}
\fntext[fn4]{Deceased}
\fntext[fn5]{Now at Bates College, Lewiston, ME 04240, USA}
\fntext[fn6]{Now at Yale University, New Haven, CT 06511, USA}
\fntext[fn7]{Also at Centre for Energy Research and Development, Obafemi Awolowo University, Ile-Ife 220005, Nigeria}
\fntext[fn8]{Now at iThemba Laboratory for Accelerator Based Sciences, Somerset West 7129, South Africa}
\fntext[fn9]{Now at Pacific Northwest National Laboratory, Richland, WA 99352, USA}

\begin{abstract}
Radon-induced backgrounds from \bi\ decays remain a significant challenge for liquid xenon (LXe) detectors searching for neutrinoless double beta decay (\nldbd{}). 
We present a simulation-based study of the \bi\ background in nEXO, incorporating a physically motivated model of the transport and implantation of radon progeny into the detector cathode. 
We show that the spatial distribution of scintillation light provides an additional method for background discrimination for events surviving standard energy, event topology, and fiducial volume selections. 
A simple discriminator based on the scintillation light detected below the cathode rejects 41\% of the remaining \bi\ backgrounds while retaining 95\% of \nldbd\ signal events, and a convolutional neural network classifier improves this rejection to 57\% at the same signal efficiency.
An updated optical model provides a more accurate estimate of the light collection efficiency below the cathode than used in previous work, increasing the expected \bi{}-\po{} (Bi-Po) coincidence-tagging efficiency for bottom-surface \bi{} decays from less than 1\% to approximately 50\%.
Together, these results demonstrate improved rejection of cathode-implanted \bi{} backgrounds in nEXO through complementary prompt scintillation-based discrimination and delayed coincidence tagging.
\end{abstract}
\end{frontmatter}

\setcounter{section}{0}
\section{Introduction}
The observed asymmetry between matter and antimatter in the universe remains one of the most profound open questions in fundamental physics. 
In leptogenesis scenarios,  
the matter-antimatter imbalance may originate from lepton-number-violating interactions in the early universe~\cite{Kuzmin:1985mm, Fukugita:1986hr}. 
Neutrinoless double beta decay (\nldbd{}) is a hypothetical nuclear process in which two neutrons decay into two protons, emitting two electrons without accompanying antineutrinos~\cite{Majorana:1937vz, Furry}, thereby violating lepton number conservation by two units ($\Delta L =2$). 
Such a process is only possible if neutrinos are Majorana fermions, i.e., identical to their own antiparticles \cite{Schechter:1981bd}. 
The search for \nldbd{} is therefore a key experimental probe for physics beyond the Standard Model, with implications for the origin of the baryon asymmetry as well as the absolute neutrino mass scale \cite{Dolinski:2019nrj, Agostini:2022zub, Adams2022}. 

Several isotopes are being pursued in a complementary experimental program to search for \nldbd{}. 
Among these, \xe{} is particularly well suited for large scale detectors that employ liquid xenon (LXe) time projection chambers (TPCs). 
LXe TPCs provide a scalable detection technology with effective self-shielding against external backgrounds, three dimensional event reconstruction, and a monolithic geometry in which the source and detector form a single homogeneous volume. 
These detectors have been widely applied in rare event searches, setting leading constraints in searches for dark matter~\cite{PandaX:2024qfu, LZ_2025, XENON:2025wimp} and \nldbd{}~\cite{Anton_2019}. 

The nEXO design extends this approach to a tonne-scale detector targeting a \nldbd{} half-life sensitivity of $T^{0\nu}_{1/2} > 10^{28}$ years \cite{Adhikari_2021}. 
Achieving this sensitivity requires stringent control and characterization of radioactive backgrounds. 
Among internal backgrounds, \rn{} and its progeny are a particularly challenging source of contamination in LXe detectors \cite{Cui:2024ltd, XENON:RnRemoval}. 
Estimates from the most recent nEXO sensitivity study indicate that radon-induced backgrounds are the dominant contribution to the background budget in the region of interest around the \xe{} \nldbd{} Q-value, $Q_{\beta \beta}$ = \qty{2458}{\keV}~\cite{Adhikari_2021}. Radon progeny backgrounds are expected to remain an important consideration for future tonne-scale liquid xenon detectors, including proposed experiments such as XLZD~\cite{XLZD:2024pdv}.

Radon, as a noble gas, can emanate from detector materials and dissolve uniformly into the active LXe volume. 
Its decay chain includes \bi{}, whose $\beta$ decay can in rare cases be accompanied by emission of a $\gamma$ with energy within \qty{10}{\keV} of $Q_{\beta \beta}$, making it a particularly problematic background for \nldbd{} searches.  
However, unlike a true \nldbd{} event, the \bi{} decay is accompanied by additional correlated signatures that can be used for background rejection if sufficient energy is deposited in active regions.
In particular, the coincident $\beta$ particle and the subsequent $\alpha$ decay of the daughter \po{} nucleus, which has a half-life of \qty{164}{\micro\second}, can each provide an independent tagging signature. 

Previous sensitivity projections for radon-induced backgrounds in nEXO~\cite{Adhikari_2021} relied on simplified, conservative estimates of backgrounds from radon progeny, particularly in the modeling of expected surface activity. 
In this work, we implement a physically motivated simulation of \bi{} implantation and its depth distribution in the nEXO cathode, as well as a detailed simulation of the energy deposits in the cathode and surrounding LXe produced during decays.
Building on this improved modeling and the highly pixelated scintillation readout provided by nEXO's barrel-mounted silicon photomultiplier (SiPM) arrays, we demonstrate enhanced rejection of \bi{} backgrounds through light-based discrimination of $\gamma$ and $\beta$ deposits as well as delayed \bi{}-\po{} (Bi-Po) coincidence tagging relative to previous nEXO sensitivity projections~\cite{Adhikari_2021}.

\section{The nEXO Detector}

\begin{figure}[t!]
    \centering
     \includegraphics[width=1.0\linewidth]{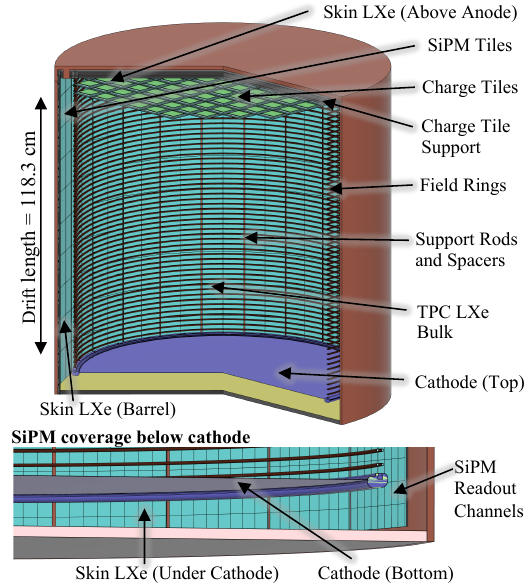}
     \caption{ 
     Schematic of detector geometry.
     Top: Full detector view showing the cylindrical field cage, charge readout tiles, and SiPM tiles arranged along the barrel.
     Bottom: Close-up of the cathode region, illustrating the extension of SiPM readout channels below the cathode.
     }
    \label{fig:detector}
\end{figure}

nEXO is a proposed \nldbd{} experiment that would deploy 5 tonnes of xenon enriched to 90\% in \xe{} in a low-background LXe TPC~\cite{nEXO:2018ylp}. 
The detector, shown schematically in Fig.~\ref{fig:detector}, is designed to measure both ionization charge and scintillation light produced by energy deposits in the active LXe volume.
Scintillation light in the \qty{175}{\nano\meter} vacuum ultraviolet band is detected by \ncount{7680} SiPM readout channels (each $3 \times 2$~cm$^2$), grouped into 480 SiPM tiles distributed along the detector barrel \cite{Gallina_2022, nEXOphotondet}.

The active region is defined by a cylindrical field cage constructed from radiopure materials. A drift electric field of \qty{400}{\volt\per\centi\meter} is generated by applying \qty{-50}{\kilo\volt} to a thin continuous copper cathode plane at the bottom and holding the charge sensitive anode plane at the top at ground. Uniformity of the drift field is maintained through the use of concentric copper field shaping rings connected by resistors that grade the potential drop between the cathode and anode \cite{nEXO:2018ylp}. 
Ionization electrons produced in the bulk LXe drift towards the anode under this field, where they are collected by \ncount{3840} charge readout channels distributed over 120 charge tiles~\cite{Jewell_2018, nEXO:2019charge, Li:2025ows}. 
The z-position of an energy deposit is reconstructed from the electron drift time, the interval between the prompt scintillation signal and the arrival of ionization electrons at the anode, while the transverse (x,y) position is reconstructed from the charge distribution across the charge tiles.

A buffer skin region of LXe surrounds the field cage. 
Ionization signals produced in this region will not be collected by the charge readout system, limiting reconstruction capabilities. 
However, the field cage is optically open \cite{nEXO:2020mec}, with SiPM tiles lining the barrel of the detector, and extending \qty{6}{\centi\meter} below the cathode as shown in the lower panel of Fig.~\ref{fig:detector}. 
As a result, scintillation light can be detected from both inside and outside the field cage volume.

Photon transport and collection play a central role in identifying scintillation signals, which is particularly important for events occurring outside the active volume where charge collection is limited.
In the previous sensitivity study \cite{Adhikari_2021}, Al+MgF$_2$ coatings were assumed to be applied to the cathode and field shaping rings to increase photon reflectivity and light collection efficiency (LCE), with a conservative 80\% reflectivity~\cite{Liu:2022hti, Bajpai:2024gep}. 
In this work, we also assume the coating is extended to include the bottom surface of the detector vessel, increasing photon collection below the cathode where no ionization signal is available.
This modification improves sensitivity to scintillation signatures associated with near cathode surface backgrounds, and is incorporated directly into the simulation framework used for background rejection.

\section{Radon Background Characteristics}

\begin{figure}[t!]
    \centering
    \includegraphics[width=1.0\linewidth]{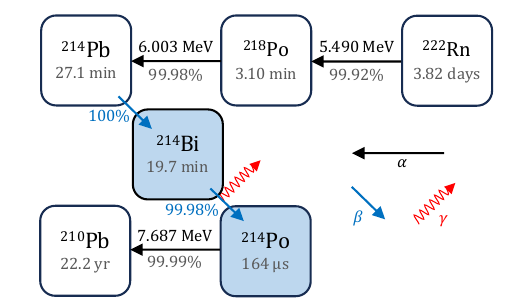}
     \caption{ Partial decay scheme of the $^{222}$Rn portion of the $^{238}$U series. 
     Branching ratios and half-lives are given for each isotope~\cite{ENSDF}.
     Isotopes used for coincidence tagging (\bi-\po{}) are highlighted in blue. 
     Only the 2448 keV $\gamma$ near $Q_{\beta\beta}$ is shown; other $\gamma$ emissions are omitted for clarity.
     }
    \label{fig:decay}
\end{figure}

Although materials for low-background rare event detectors are selected for low intrinsic activity, residual contaminants from the uranium and thorium decay chains remain. In particular, $^{222}$Rn, a radioactive noble gas with a half-life of 3.8 days in the $^{238}$U decay chain presents a critical challenge. This lifetime is sufficient for the gas to emanate from surrounding materials and distribute throughout the detector volume before decaying. Its subsequent decay chain is shown in Fig. \ref{fig:decay}.

Of particular concern is the decay of \bi{}. \bi{} decays to \po{} through $\beta$ emission and has a ground-state to ground-state Q-value of \qty{3269}{\keV}. A small fraction of decays (branching ratio of 1.5\%) produce $\gamma$-rays with energy at \qty{2448}{\keV}~\cite{Zhu_214decays:2021qss}. This is within \qty{10}{\keV} of the \xe{} \qbb{} and in a LXe TPC cannot be distinguished from a \nldbd{} signal based on energy resolution alone. Within the fiducial volume of the TPC a significant fraction of the \bi\ $\gamma$ background is reduced by the self-shielding of the liquid xenon and by removing multi-scatter interactions. Nevertheless, the small rate of \qty{2448}{\keV} \bi{} $\gamma$-rays that make it unattenuated into the fiducial volume and deposit their entire energy in a single photoelectric interaction are the dominant source of background for \nldbd\ searches.

A range of mitigation strategies has been developed to reduce radon backgrounds in LXe detectors, including material screening and selection to suppress intrinsic radioactivity and limit radon emanation \cite{Hoppe:2014nva, Albert:2015nta, LZ:2020radiopurity, XENON:2021radiopurity, nEXO:Anker, MacLellan:nEXOradioassay}, as well as active removal techniques such as cryogenic distillation and adsorption based filtration \cite{ Cui:2024ltd, XENON:RnRemoval, Pushkin:2018wdl}. In this work we focus on software analysis cuts to identify the \bi{} $\gamma$-ray background events through the detection of coincident energy deposits.

The \SI{2448}{\keV} $\gamma$-ray is accompanied by a coincident $\beta$ particle emission ($Q_\beta$ = 3269-2448 = \qty{821}{\keV}) that deposits additional energy near the decay location. The \bi{} $\beta$ decay is in turn quickly followed by the $\alpha$ decay of \po{} to $^{210}$Pb ($Q_\alpha = \qty{7833}{\keV}, T_{1/2} = \qty{164}{\micro\second}$). If these coincident energy deposits occur in an active volume and are detectable, they can be used to tag and remove \bi{} background events. In the following sections we will describe our efforts to accurately simulate these coincident energy deposits and their corresponding detectable light and charge signals in order to estimate the overall background rejection efficiency.

\section{Tracking Radon Progeny}
Unlike $^{222}$Rn, its progeny are chemically reactive and frequently produced in a positively charged state \cite{EXO-200:2015ura}. When these decays occur in the active xenon volume, the charged progeny drift along electric field lines and can plate out on cathodic detector surfaces. In the nEXO design, Rn daughters plated out on surfaces contribute approximately 50\% of the total background budget~\cite{Adhikari_2021}.

Radon progeny transport was modeled by propagating the daughters of the $^{222}$Rn decay chain under the influence of the detector electric field. 
In this study, initial $^{222}$Rn atoms are assumed to be neutral and uniformly distributed throughout the LXe, although this need not hold in a large detector with limited mixing~\cite{LZ:2025xxf}. 
After each decay, the daughter nucleus is assigned a charge state based on measurements of the charge state of radon progeny in EXO-200~\cite{EXO-200:2015ura}: i.e., $50.3 \pm 3.0\%$ of daughters following $\alpha$ decays and $76.3 \pm 6.2\%$ following $\beta$ decays are assumed to be ionized.
These values were measured for the $^{222}$Rn $\alpha$ and $^{214}$Pb $\beta$ decays respectively and are applied to all decays in the chain by decay type, an extrapolation supported for the $\alpha$ case in Ref.~\cite{EXO-200:2015ura}.
Charged progeny drift along the electric field lines, while neutral atoms remain stationary and decay in place. The charge state is assumed to remain fixed between decays, and charged progeny are transported using the high-purity ion mobility of \SI[inter-unit-product=\ensuremath{{}\cdot{}}]{0.39}{\centi\meter\squared\per\kilo\volt\per\second}. 
The reduction in ion mobility observed during drift~\cite{EXO-200:2015ura} is neglected, which is conservative
since it would lower the fraction of progeny reaching the cathode.
If a charged particle reaches a detector surface, it is assumed to stick to the surface~\cite{Chambers2023}, with the possibility of implantation (see Section~\ref{sec:implantation}).
Uncertainties in both the ion mobility and ionization fractions are propagated in the simulation.

In practice, radon progeny transport will also be governed by advection due to LXe fluid flow. If the fluid flow velocity is comparable to or larger than the ion drift velocity in the applied electric field, the trajectories of radon progeny could be significantly altered. It is difficult to accurately model the LXe fluid flow at the design phase of the experiment, but the flow patterns in an operating TPC can be deduced from the reconstructed positions of fast-coincidence isotopes~\cite{EXO-200:2015ura, LZ:2025xxf, XENON:2024lbh}.  In addition, the probability that charged or neutral daughters adhere to detector surfaces following transport is not experimentally well constrained and may depend on local detector conditions. The present model neglects these considerations, which may affect the predicted distribution of radon progeny in the detector.

\begin{table}[t!]
    \centering
    \begin{adjustbox}{max width=1.0\linewidth}
    \setlength{\tabcolsep}{20pt}
    \begin{tabular}{l|l}
        \hline
        Decay Location $\ell$ & $f_{\ell}$  \\
        \hline
        Cathode top surface      & 0.60  \\
        Cathode bottom surface   & 0.06 \\
        Field rings surfaces     & 0.19 \\
        TPC LXe bulk             & 0.13  \\
        Skin LXe barrel          & 0.01 \\
        Skin LXe above anode     & 0.01  \\
        Skin LXe under cathode   & 0.00  \\
        \hline
    \end{tabular}
    \end{adjustbox}
    \caption{Predicted spatial distribution of \bi{} decays from the radon progeny transport simulation, given as the decay fraction $f_{\ell}$ at each decay location $\ell$.}
    \label{tab:decay_table}
\end{table}

In the absence of these effects, the distribution of $^{222}$Rn progeny is determined by the combination of radioactive decay and electric field drift.
Using the detailed design of the nEXO detector geometry, we evaluate the resulting spatial distribution of \bi{} decays.
The fractional decay rates on detector surfaces are summarized in Table \ref{tab:decay_table}, and are in good agreement with the prior simplified estimates~\cite{Adhikari_2021}. 
Small differences are calculated for the cathode surfaces, including a decrease in the top surface fraction from 0.62 to 0.60, and an increase on the bottom surface from 0.04 to 0.06. These changes reflect a shift in the cathode position in the nEXO design which modestly increased the LXe volume below the cathode relative to the configuration assumed in Ref.~\cite{Adhikari_2021}. 

As shown in Table~\ref{tab:decay_table}, a large fraction of the \bi{} decays from radon within the LXe occurs on the negatively biased cathode, and these decays are expected to contribute approximately 33\% of the total background budget of the nEXO experiment~\cite{Adhikari_2021}. In the remaining sections we therefore focus on the decays on the cathode, though a similar approach could also be applied to decays on the field ring surfaces.

\section{Implantation}
\label{sec:implantation}

\begin{figure}[t!]
    \centering
    \includegraphics[width=0.98\linewidth]{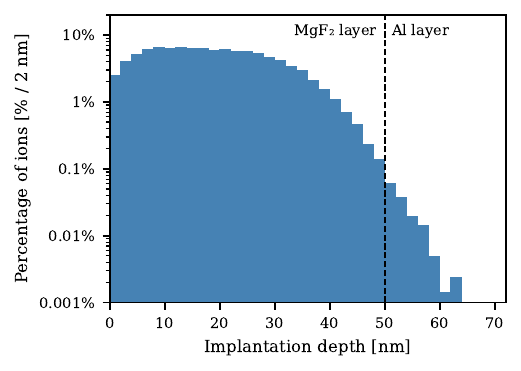}
     \caption{ 
     The simulated distribution of \bi{} atoms within the cathode layers after implantation.
     The mean implantation depth is 19~nm.
     Less than 0.14\% of atoms are implanted at depths greater than \qty{50}{\nano\meter} (within the reflective aluminum layer).
     }
    \label{fig:implant}
\end{figure}

As seen in Fig.~\ref{fig:decay}, the \rn{} decay chain first decays through \isotope{Po}{218} and \isotope{Pb}{214} before reaching the main background contributor within the TPC, \bi{}.
If \isotope{Po}{218} settles on a surface prior to decaying, the recoil from its
$\alpha$ decay can implant the nucleus at some depth into the surface material or, conversely, eject it from the surface.
To quantify these effects, the Monte Carlo (MC) simulation package SRIM (Stopping and Range of Ions in Matter) was used~\cite{ziegler2010srim}.

Implantation depths are simulated with SRIM using a model of the cathode surfaces that includes the reflective and protective coatings applied to both faces of the copper. 
From the outer surface inward, each face consists of \qty{50}{\nano\meter} of MgF$_2$, \qty{100}{\nano\meter} of Al, and another \qty{10}{\nano\meter} of MgF$_2$, followed by the \qty{0.127}{\milli\meter}-thick bulk copper volume~\cite{Liu:2022hti}.

Simulations start with a \isotope{Po}{218} atom resting on the outermost surface of the MgF$_2$ layer.
The atom $\alpha$-decays to \isotope{Pb}{214}, with a nuclear recoil energy of \qty{112}{keV}.
The recoil is tracked through the cathode layers and the implantation depth is recorded. 
Recoils are simulated at multiple incident angles and weighted to produce an isotropic angular distribution, with ions recoiling into the LXe assumed to drift back to the cathode under the electric field~\cite{Chambers2023}.
The atom then $\beta$-decays into \bi{} at the corresponding depth, and the additional displacement is added to find the final implantation depth within the cathode, which is shown in Fig.~\ref{fig:implant}.

Less than 0.14\% of \bi{} atoms reach the aluminum layer, with fewer
than 0.005\% exceeding a depth of \qty{60}{\nano\meter}.
The distribution of \bi{} atoms has a mean implantation depth of \qty{18.8}{\nano\meter}, with a higher density toward the outer surface of the MgF$_2$ layer.
This implantation is used to determine the energy deposits within the LXe created by \bi{} decays in the following section, accounting for energy that may be lost when exiting the cathode surface layer prior to reaching the LXe.

\section{Event Simulations}
\label{sec:simulations}
\subsection{\bi{} Event Simulation}

\begin{figure}[t!]
    \centering
    \includegraphics[width=0.98\linewidth]{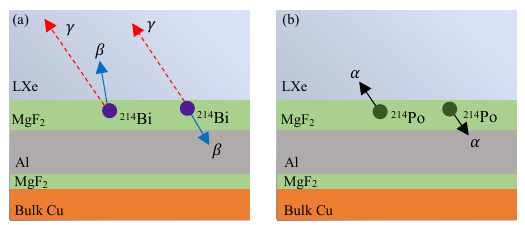}
     \caption{ 
      Schematic of (a)\bi{} and (b)\po{} decays for nuclei implanted near the surface layers of the cathode. The topology shown is representative of both the top and bottom cathode surfaces. In each panel the left case shows the emitted charged particle directed into the LXe, where it can produce a detectable signal, while the right shows it directed into the cathode material. Dimensions not to scale.
     }
    \label{fig:decay_cartoon}
\end{figure}

The \bi\ \SI{2448}{\keV} $\gamma$-ray has an attenuation length of \SI{8.7}{\cm} in liquid xenon and can travel a significant distance before interacting, while the coincident $\beta$ particle, with a kinetic energy of up to \SI{821}{\keV}, has a maximum range of \SI{0.19}{\cm}, depositing energy near the decay vertex (Fig. \ref{fig:decay_cartoon}). The \bi{} decay can therefore produce interactions with a measurable spatial separation, particularly along the $z$ direction, observable through reconstruction of the separated charge deposits or from the position separation encoded in the photon hit pattern on the individual SiPM channels.

To evaluate the ability to detect these coincident $\beta$ deposits, MC simulations of the \bi{} decay and the corresponding nEXO detector response were implemented using the Geant4 simulation toolkit~\cite{AGOSTINELLI2003250,1610988,ALLISON2016186} in combination with a custom version of the Noble Element Simulation Technique (NEST) package~\cite{Szydagis:2011tk,nest2.0.1} to calculate the production of electrons and scintillation photons.
These quantities, together with their generated positions, are stored as individual ``hits'' which are subsequently propagated through the detector response model.
A detailed description of the geometry and construction of the nEXO detector model and response can be found in Refs.~\cite{Adhikari_2021, PhysRevC.97.065503}.
To accurately simulate \bi{} atoms implanted within the cathode surface, the previously mentioned layers of MgF$_2$ and Al were also implemented on the surfaces of the Geant4 model for the cathode volume. 

\begin{figure}[t!]
    \centering
     \includegraphics[width=0.98\linewidth]{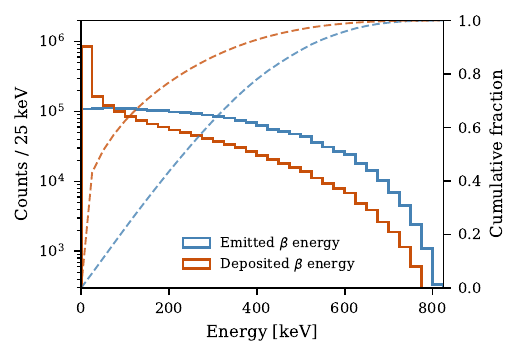}
     \caption{ 
     Kinetic energy spectrum of the primary $\beta$ particles emitted in coincidence with the 2448~keV $\gamma$ branch of \bi{} (blue), compared with the energy those $\beta$ particles deposit in the LXe when originating from \bi{} implanted in the cathode (orange). Dashed curves show the corresponding cumulative fractions (right axis).
     }
    \label{fig:betaEGen}
\end{figure}

Due to the need for high statistics for validation and the low branching ratio of signal-like \bi{} decays, a custom particle generating action was written for these events, which specifically simulates a \qty{2448}{keV} $\gamma$ in combination with a $\beta$ particle.
The $\beta$ particle energy (shown in Fig.~\ref{fig:betaEGen}) is determined by randomly sampling from a generator-level distribution of primary $\beta$ energies obtained from a separate dedicated simulation of \bi{} decays, recording all $\beta$ particles emitted in coincidence with the \qty{2448}{keV} $\gamma$.
This method includes the small fraction of events where the \qty{2448}{keV} $\gamma$ is part of a cascade, reducing the energy available for the $\beta$.

A total of $5 \times 10^{6}$ $\gamma$-$\beta$ event pairs were generated with the starting location within the cathode surface according to the implantation distribution shown in Fig.~\ref{fig:implant}, with $2.5 \times 10^{6}$ events for each configuration (top and bottom).
The particles are generated in pairs, each with an independent, uniformly distributed angular distribution.
Their interactions in LXe are simulated, and only events depositing at least \qty{700}{keV} of total energy in LXe are retained, leaving \num{2.1e6} events for each
implantation configuration.

Ionization and scintillation production is then calculated from the energy of the interactions within the LXe on a per-hit basis, using electric field values at the region in which the interaction occurs~\cite{Adhikari_2021}.
Within the active TPC region, the drift field is \qty{400}{\volt\per\centi\meter}, consistent with the nominal nEXO operating conditions.
Outside the field cage, electric fields are region dependent, with a value of \qty{6.7}{\kilo\volt\per\centi\meter} below the cathode surface.

\begin{figure}[t!]
    \centering
     \includegraphics[width=0.98\linewidth]{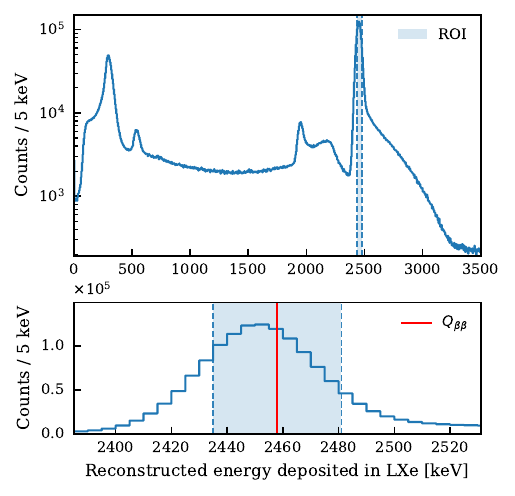}
     \caption{Reconstructed energy spectrum in the LXe for $\gamma$–$\beta$ event pairs. The upper panel shows the full spectrum on a logarithmic scale, while the lower panel provides a linear-scale zoom of the ROI defined around $Q_{\beta\beta} = 2458$ keV (red line).
     The main peak at 2448~keV comes from $\gamma$ interactions that fully deposit their energy within the LXe, with the additional events coincident with a $\beta$ depositing energy in the LXe appearing as a tail at higher energy, extending to the $Q$-value of the decay of 3269~keV.
     At lower energies, a broad Compton continuum and discrete structures from pair-production escape peaks and annihilation-photon interactions arise from $\gamma$ interactions that only partially deposit energy before escaping the detector.
     }
    \label{fig:Bi_edep}
\end{figure}

Events are subsequently propagated through the detector response~\cite{Adhikari_2021} and used to evaluate discrimination performance using charge and light based tagging methods.
Fig.~\ref{fig:Bi_edep} shows the resulting spectrum of total reconstructed energy deposition per event.
The region of interest (ROI) is defined as the full width at half maximum (FWHM) of a peak centered at $Q_{\beta\beta}$. Using the projected nEXO energy resolution of $\sigma/E = 0.8\%$ at $Q_{\beta\beta}$~\cite{Adhikari_2021}, the ROI spans [2435, 2481]~keV.

\subsection{\po{} $\alpha$ Decay 
  Simulation}
The existing nEXO analysis incorporates tagging of the \qty{7.7}{MeV} $\alpha$ decay from \po{}~\cite{Adhikari_2021, PhysRevC.97.065503}, which follows the preceding \bi{} decay with a half-life of \qty{164}{\micro\second}.
The large energy deposited by the $\alpha$ particle makes these decays readily identifiable when they occur within the active LXe volume~\cite{nEXO:2020mec}.
By searching for a delayed $\alpha$ signal following a signal-like event, radon induced backgrounds originating in the bulk LXe can be efficiently vetoed. 
A finite veto window of \qty{1.5}{\milli\second} results in only 0.18\% of \po{} $\alpha$ decays remaining untagged due to the decay occurring outside the veto period.

\begin{figure*}[t!]
    \centering
     \includegraphics[width=1.0\linewidth]{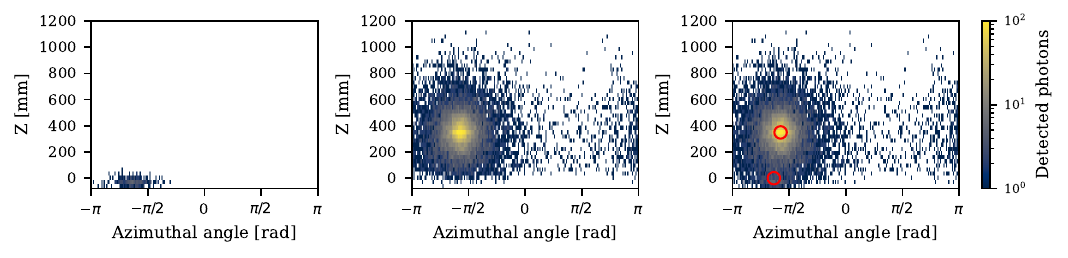}
     \caption{ 
     Simulated SiPM photon hit patterns for a \bi{} event implanted on the bottom of the cathode and with a reconstructed energy in the ROI. In this case, a 604 keV $\beta$ that survives all reconstruction and DNN cuts also interacts near the cathode, although the reconstructed position of the event lies within the inner 2 tonne fiducial volume. 
     The detector is shown in an unwrapped cylindrical projection, where each pixel corresponds to a $3 \times 2$~cm$^2$ light readout channel.
     (left) $\beta$ induced cluster exhibiting a localized light distribution near the cathode.
     (middle) $\gamma$ induced cluster that produces a more diffuse and displaced hit pattern. 
     (right) Combined clusters illustrating the spatial separation between $\gamma$ and $\beta$. Red circles indicate the projected coordinates of the $\gamma$ and $\beta$ interactions in the simulation.
    }
    \label{fig:bi_hitpattern}
\end{figure*}

The effectiveness of this tagging technique is reduced for decays occurring on detector surfaces.
In these cases, the emitted $\alpha$ particle may be directed into detector materials rather than into the LXe (Fig.~\ref{fig:decay_cartoon}), resulting in little or no observable energy deposition in the active volume.
MC simulations of \po{} decays at the cathode surface show a strongly bimodal distribution of energy deposits in LXe, with events clustering near \qty{0}{MeV} when the $\alpha$ is emitted into the cathode and near the full $\alpha$ kinetic energy when emitted into LXe. 
For top-surface decays, $\alpha$ particles directed into the drift LXe are already incorporated in the existing nEXO sensitivity analysis~\cite{Adhikari_2021}.
For bottom-surface decays, the simulated fraction of $\alpha$ decays that deposit energy in the LXe is used to estimate the delayed-$\alpha$ tagging efficiency in Section~8.4.
Scintillation yields for $\alpha$ particles are estimated directly from NEST~\cite{nestpy}.
Since the $\alpha$ light yield is essentially field-independent above $\sim$\qty{100}{\volt\per\centi\meter}~\cite{Aprile:2009dj}, we use the NEST yield at \qty{7}{\kilo\volt\per\centi\meter} in the region below the cathode.

\section{Tagging Methodology}
\label{sec:taggingmethod}

In simulation, the average energy deposited in the LXe per emitted \bi{} $\beta$ particle is approximately \qty{116}{keV} (see Fig.~\ref{fig:betaEGen}), producing an average of \ncount{4900} scintillation photons and \ncount{5300} ionization electrons in the \qty{400}{\volt\per\cm} drift region above the cathode~\cite{Szydagis:2011tk}.
For $\beta$ particles entering the skin LXe below the cathode, where charge is not collected, the average scintillation signal is approximately \ncount{1500} photons, as the higher local field of \qty{6.7}{\kilo\volt\per\cm} strongly suppresses recombination. 
The delayed \po{} $\alpha$ particle deposits $\sim$\qty{7.7}{MeV} in LXe, producing approximately $9\times10^5$ photons in the same region~\cite{nestpy}. 
These charge and scintillation observables form the basis of the background discrimination techniques described in the following sections.
The evaluation of charge-based reconstruction, scintillation-based $\beta$ tagging, and delayed Bi-Po tagging requires distinct event selections and reconstruction procedures tailored to each method. 

\subsection{\bi{} Tagging Methodology}
\subsubsection{Charge Signal Reconstruction}
\label{sec:chargerecon}
The charge-based reconstruction analysis quantifies the rejection of cathode \bi{} backgrounds provided by the standard nEXO reconstruction framework. 
In particular, it evaluates how often a \bi{} event with a signal-like $\gamma$ deposition is rejected due to the reconstructed charge signal from the accompanying $\beta$.
Events of interest are first selected by requiring the true energy deposited within the fiducial volume to fall within the ROI.
The fiducial volume is defined as the active LXe region at least \qty{2}{\cm} from all detector surfaces. The coincident $\beta$ deposits at the cathode surface, outside this volume, and
therefore does not contribute to this sum.                  

Charge reconstruction then follows the standard nEXO framework \cite{Adhikari_2021}.
The standoff distance, defined as the minimum distance between a reconstructed charge cluster and the nearest detector surface, provides discrimination against backgrounds originating from detector surfaces.
Electronic noise is modeled by including a noise level for each charge waveform with $\sigma = 70$~electrons~\cite{Adhikari_2021}, which in turn sets the minimum threshold for reconstructing charge deposits. 
This noise level is approximately a factor of two smaller than the nEXO requirement~\cite{Li:2025ows}, representing an optimistic level that could be reached in the final detector, provided optimal electronics performance.

To isolate the impact of the coincident $\beta$ on charge reconstruction, the $\beta$ and $\gamma$ energy depositions are then separated using truth-level Geant4 generator information. All deposits within a \qty{5}{\milli\meter} radius of the \bi{} generator position are attributed to the $\beta$; those outside are attributed to the $\gamma$. Note that because events are selected based on the $\gamma$ energy deposited within the fiducial volume, $\gamma$ interactions near the cathode are rare by construction. Radii of 2–\qty{5}{\milli\meter} give consistent mean inferred $\beta$ energies to within 1\%, with \qty{5}{\milli\meter} chosen for marginally better bremsstrahlung containment. 
This inferred $\beta$ energy is used to evaluate the standoff distance efficiency in Section~\ref{sec:chargebeta}.

\subsubsection{Light Signal Reconstruction}
\label{sec:lightrecon}
The light-based analysis focuses on events that pass all standard reconstruction cuts, including energy, standoff, light-to-charge ratio, and DNN-based topology~\cite{Adhikari_2021}. 
These events constitute the \bi{} background population that are not distinguished from signal.
The energy distribution for these events prior to the application of selection cuts is shown in the lower panel of Fig.~\ref{fig:Bi_edep}.
 
Energy depositions generated with Geant4 are converted to scintillation photon yields using NEST, as described in the previous section. 
The resulting NEST energy depositions are grouped into clusters using single-linkage hierarchical clustering with a \qty{3}{\milli\meter} distance threshold, combining nearby deposits into distinct energy-deposition sites.
Each cluster is passed to Chroma~\cite{Seibert2011FastOM,chroma-git}, a GPU-enabled optical photon transport framework, as an isotropic point source at its energy-weighted centroid position, emitting a number of photons equal to the total NEST photon count for that cluster.
Chroma then propagates these photons through the nEXO detector geometry, modeling transport, reflections, and detection at the SiPM plane. The optical properties follow Ref.~\cite{Adhikari_2021} with modifications described in Section~\ref{sec:bipo}.

\begin{figure}[t!]
    \centering
     \includegraphics[width=1.0\linewidth]{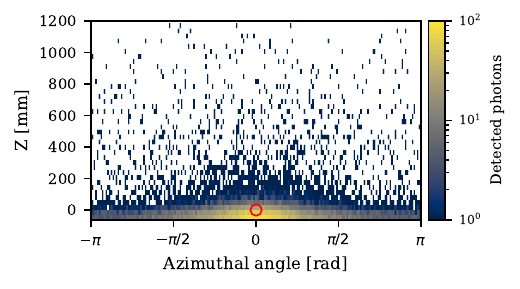}
     \caption{ 
     Simulated SiPM photon hit pattern for an $\alpha$ particle from $^{214}$Po decay that deposits 2 MeV of energy in the skin LXe below the cathode.
     The detector is shown in an unwrapped cylindrical projection, with  $3 \times 2$~cm$^2$ light readout channels.
     The light distribution is strongly localized near the cathode, reflecting the short range and energetic nature of the $\alpha$ particle. The red circle indicates the projected coordinates of the $\alpha$ interaction in the simulation.
    }
    \label{fig:alpha_light}
\end{figure}

Photon contributions from all clusters within an event are summed to produce the final simulated SiPM response, represented as a spatial hit pattern of detected photon counts across the readout channels. 
Examples of these hit patterns are shown in Fig.~\ref{fig:bi_hitpattern} and Fig.~\ref{fig:alpha_light}. 
The detector is represented in an unwrapped cylindrical projection, where each pixel corresponds to a $3\times2$~cm$^2$ light readout channel. 
Detected photon counts include the effects of the photon detection efficiency (PDE) and any inactive regions between tiles. 
Dark noise is negligible~\cite{Gallina_2022}, and does not significantly affect the analysis for the events considered here.

This analysis is restricted to the inner 2T fiducial region (central \qty{2000}{\kilo\gram} of active LXe), which previous nEXO sensitivity studies found corresponds to the optimal fiducial volume in a simplified cut-based analysis, retaining approximately 95\% of the sensitivity of a fit to the full event distribution~\cite{Adhikari_2021}.
The lower $z$ boundary of this region is \qty{10.4}{\centi\meter} away from the cathode surface.
While this volume selection removes the majority of surface adjacent backgrounds, additional discrimination techniques are needed to further reject events without reconstructed energy deposits near the surfaces.
After the full detector response cuts and the inner 2T cut, only 926 ROI-like background events (630 bottom-implantation and 296 top-implantation) remain from the initial simulation of $5\times 10^6$ events. 

To increase statistical power for studying the rare event topologies that can evade the traditional nEXO analysis cuts, an augmented sample is constructed from the signal-like background events prior to the inner 2T cut, resulting in \ncount{4903} bottom implantation and \ncount{3139} top implantation events. 
For each event, the spatial coordinates of the $\gamma$ energy depositions are translated along the $z$-axis, with their distance beyond the inner 2T boundary sampled from an exponential distribution. This approximates the projected $z$-distribution of isotropically emitted $\gamma$-rays, with a decay constant of \qty{6.9}{\centi\meter} chosen to match the attenuation profile at that boundary.
Event topology, deposited energies, and relative spatial correlations between interaction sites are preserved. 
Optical photon propagation simulations are then performed for this higher statistics augmented sample using Chroma. 
The augmented sample is used only for training, while the 926 original simulated events remaining after all cuts are used to evaluate the performance of light-based tagging strategies in Section~\ref{sec:lightbeta}.

\subsection{Bi-Po Tagging Methodology}
\label{sec:bipo}
To evaluate the efficiency of delayed Bi-Po tagging for cathode-surface decays, the \po{} $\alpha$ scintillation signal is simulated.
Decay positions are sampled uniformly across the cathode surface.
Scintillation yields are estimated using NEST~\cite{nestpy} at the local electric field below the cathode, and the resulting photons are propagated through the detector using the Chroma optical simulation framework.
Because $\alpha$ particles deposit essentially all of their energy over sub-millimeter length scales in LXe, detailed Geant4 tracking of individual interactions is not required.

The estimated tagging efficiency depends critically on the light collection efficiency (LCE) below the cathode.
The LCE is defined as the product of photon transport efficiency (PTE) and the probability that a photon absorbed by the SiPM creates a detectable avalanche~\cite{Adhikari_2021}, which is determined from measurements of the PDE for prototype nEXO SiPMs. 
Previous nEXO sensitivity studies assumed a conservative LCE of 0.1\% in this region, resulting in an estimated tagging efficiency below 1\% for cathode-surface decays~\cite{Adhikari_2021,nEXO:2020mec}.
The updated framework incorporates several improvements to the LCE model.
Recent measurements of SiPM prototype devices for nEXO indicate an expected PDE of 21\% at 3~V overvoltage~\cite{Gallina_2022}. 
In parallel, refinements to the SiPM tile geometry increase the effective fill factor of the active regions of the photodetector system \cite{SiPM_tiles_Tsang}.
Furthermore, updated optical simulations with Chroma, including the addition of a reflective Al+MgF$_2$ layer to the bottom surface of the detector, yield an average PTE of 10.5\% in the region below the cathode. 
Taken together, these updates result in an LCE of 3.2\% in the region below the cathode, over an order of magnitude higher than previous assumptions~\cite{Adhikari_2021, nEXO:2020mec}.

\section{Tagging Performance}
\label{sec:taggingperformance}
In this section, the rejection efficiency of cathode-implanted \bi{} backgrounds is evaluated using the complementary signatures described in Section~\ref{sec:taggingmethod}.
To understand the baseline charge rejection due to $\beta$ particle energy deposits before introducing additional light-based discriminants, the standoff distance cut is evaluated as a function of $\beta$ energy and implantation location.
For signal-like background events in the inner 2T, the spatial distribution of scintillation light provides further background discrimination. This is evaluated using a simple photon count from the SiPM read-out channels forming the bottom row, which span 4 to \qty{6}{\centi\meter} below the cathode (hereafter referred to as the bottom-row SiPM channels; Fig.~\ref{fig:detector}), and a CNN-based classifier. 
Finally, the scintillation produced from the delayed $\alpha$ decay of \po{} enables identification of \bi{} decays through Bi-Po coincidences, independent of the $\beta$ and $\gamma$ signatures.

\subsection{Charge Based $\beta$-Tagging}
\label{sec:chargebeta}
\begin{figure}[t!]
    \centering
     \includegraphics[width=0.98\linewidth]{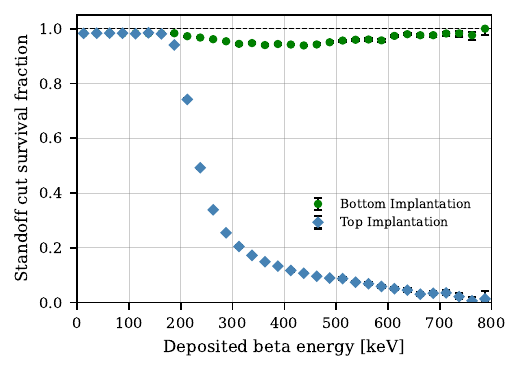}
     \caption{ 
     Standoff cut survival fraction for simulated \bi{} events with $\gamma$ energy in the ROI [2435-2481 keV] as a function of deposited coincident $\beta$ energy for bottom (green) and top (blue) cathode implantation events.
     }
    \label{fig:standoff}
\end{figure}

The standoff distance cut rejects events with reconstructed $z$ positions near the cathode surface, where radon progeny are implanted. 
To quantify the performance of this standoff cut, its rejection is evaluated on the sample with MC truth energy within the ROI described in Section~\ref{sec:chargerecon}, with no additional reconstruction cuts applied. 
As shown in Fig.~\ref{fig:standoff}, the efficiency of this cut depends strongly on the implantation location and, for top-implantation events, on $\beta$ energy.

For bottom-implantation events, the $\beta$ deposits its energy predominantly within the cathode itself or in the skin LXe below the cathode, where ionization charge is not collected.
Therefore, the survival fraction remains above $\sim$94\% across all $\beta$ energies. 
The shallow minimum near $400$~keV arises from the small fraction of $\beta$s
energetic enough to traverse the $0.127$~mm cathode and deposit charge on
the drift side, close to the cathode surface.
Away from this minimum, the small rejected population is dominated by $\gamma$ interactions near detector surfaces, predominantly Compton scatters close to the field-cage wall that produce reconstructed charge clusters within 2.0 cm of the barrel. 
These events satisfy the fiducial-volume energy selection because only a small fraction of the $\gamma$ energy is deposited near the surface, with the remainder deposited in the detector interior.

For top-implantation events, $\beta$ particles that exit the cathode deposit charge in the active drift region. 
At low $\beta$ energies, the survival fraction is similar to that of bottom-implantation events and is dominated by the same $\gamma$-induced surface interactions.
The survival fraction decreases sharply above $\sim$\qty{200}{keV} as $\beta$ deposits near the cathode begin to exceed the charge reconstruction threshold, with 92\% of events rejected above \qty{400}{keV}. 
Integrated over the entire deposited energy spectrum (see Fig.~\ref{fig:betaEGen}), approximately 18\% of top-implantation events are rejected.
The surviving population in both implantation configurations motivates the use of the light channel to further reject background events that can evade the charge-based cuts alone.

We note that additional charge tagging methods can be employed to improve rejection of low energy $\beta$ events at the top of the cathode. The charge from these events always occurs a full TPC drift time after the light signal from the coincident $\gamma$ interaction. For every potential \nldbd\ signal in the ROI one could therefore search for a charge signal in the narrow time window corresponding to an event at the cathode, allowing for sensitivity to signals below the typical noise thresholds. The sensitivity of such a search can be further improved if one were able to localize the potential location of such coincident $\beta$s in the x-y plane. This can be done by looking for $\alpha$ decays at the cathode prior to the potential \nldbd\ signal. Recall that \bi\ is implanted in the cathode if the parent \isotope{Po}{218} reached the cathode and the subsequent $\alpha$ decay drives the recoiling nucleus into the surface while the $\alpha$ is emitted into the LXe. The charge and light signal from the \isotope{Po}{218} $\alpha$ (\qty{6.00}{\MeV}) should readily allow for x-y position reconstruction of the event. For every potential \nldbd\ signal one could therefore search for cathode $\alpha$ events prior to the signal and use the x-y reconstructed positions as candidate locations for performing a more sensitive coincident \bi\ $\beta$ search. Given the half-lives of \isotope{Pb}{214} (\qty{27.1}{min}) and \bi\ (\qty{19.7}{min}), a search in the prior \qty{3}{hours} would have an efficiency of 97\%. With the nEXO target of less than 600 atoms of \rn\ in the LXe \cite{nEXO:2018ylp}, we expect roughly 8 \isotope{Po}{218} $\alpha$ decays on the cathode during that period, greatly narrowing the potential x-y locations. The planned use of radiopure electroformed copper for the cathode material makes accidental $\alpha$ decays from intrinsic radiocontamination in the cathode negligible~\cite{MacLellan:nEXOradioassay}. 

\subsection{Bottom-Row Scintillation Tagging}
\label{sec:lightbeta}
The discriminating power of scintillation light is evaluated using the original simulated \bi{} events that survive all standard nEXO reconstruction cuts within the inner 2T fiducial volume (Section~\ref{sec:lightrecon}).
As illustrated in Fig. \ref{fig:bi_hitpattern}, $\beta$ energy depositions occurring near the cathode produce localized light signals on the bottom-row SiPM channels, while $\gamma$ interactions produce more diffuse and displaced patterns that can mimic \nldbd\ signals within the inner 2T fiducial volume. 
Prior nEXO simulation studies considered only the ratio of the total scintillation to
the measured ionization signals~\cite{Adhikari_2021, nEXO:2020mec}. 

\begin{figure}[t!]
    \centering
    \includegraphics[width=1.0\linewidth]{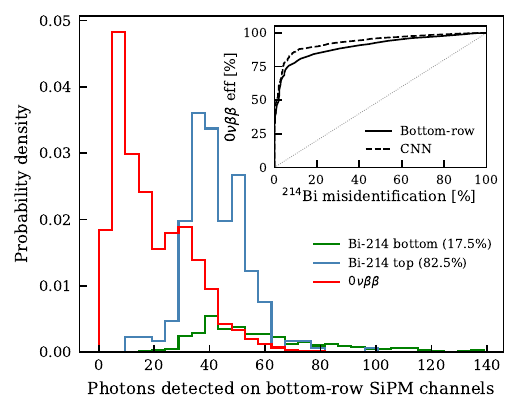}
     \caption{ 
     Probability density of photons detected on the bottom-row SiPM channels for inner 2T \bi{} and $0\nu\beta\beta$ signal (red) events.
     The \bi{} top (blue) and bottom (green) implantation populations are scaled by their relative contributions to the \bi{} background model, such that the combined \bi{} background distribution integrates to unity, while the $0\nu\beta\beta$ signal distribution is independently normalized to unit area.
     Inset: ROC curves for the weighted combined \bi{} background for the bottom-row photon count discriminant (solid) and CNN (dashed). 
    }
    \label{fig:light_cut}
\end{figure}

Using the bottom-row SiPM photon count as a discriminating observable, light-based discrimination performance is evaluated by comparing a $0\nu\beta\beta$ signal population that is uniformly distributed within the inner 2T to signal-like background events in this volume. 
Fig. \ref{fig:light_cut} shows the probability density of bottom-row SiPM photon counts for the \nldbd\ signal and the weighted combined \bi{} background.
The bottom-row SiPM photon count includes scintillation from both the $\beta$ and $\gamma$ interactions, and therefore reflects contributions from cathode-adjacent $\beta$ deposition as well as the $z$-dependent light collection from $\gamma$ interactions within the inner 2T. 
Top-implantation decays are substantially more abundant in the transport simulation
(Table~\ref{tab:decay_table}), and thus constitute 82.5\% of the weighted signal-like
background population within the inner 2T, despite being rejected more efficiently by the
standard reconstruction and fiducial selections. The combined discriminator performance is therefore largely set by this population.
The corresponding receiver operating characteristic (ROC) curve for the combined background is shown in the inset of Fig. \ref{fig:light_cut}.

nEXO employs a multiparameter likelihood fit to optimally combine reconstructed event observables~\cite{Adhikari_2021}. For illustration, we instead evaluate the performance of simple threshold cuts on the bottom-row SiPM photon count, selecting a threshold at a given signal efficiency and determining the corresponding background rejection.  
For example, retaining 95\% of \nldbd\ signal events within the inner 2T fiducial volume would allow up to 41\% of the remaining cathode \bi{} backgrounds present in previous nEXO sensitivity evaluations to be rejected, with minimal loss of signal. 
At an operating point near the steepest region of the ROC curve, 98\% of these backgrounds are rejected while retaining 49\% signal efficiency.

\subsection{CNN-Based Scintillation Tagging}
To evaluate whether additional discrimination can be extracted from the full scintillation-light distribution, a convolutional neural network (CNN) was trained on the SiPM hit patterns to distinguish signal and background events.
The network consists of three convolutional layers with 16, 32, and 64 filters, trained using a binary cross-entropy loss on log-scaled hit patterns. 
Because the augmented sample of signal-like background events in the inner 2T is limited ($\sim$7{,}000 unique events), rotations in $\phi$ are applied during training, exploiting the azimuthal symmetry of the SiPM array. 
Each training event is included at four fixed rotations, with additional random rotations applied as events are loaded.
Given the limited number of unique events, a relatively simple CNN architecture is adopted. 
The resulting ROC curve is also shown in the inset of Fig. \ref{fig:light_cut}. 
At 95\% signal efficiency, the CNN rejects 57\% of \bi{} backgrounds in the original inner 2T sample, compared to 41\% for the bottom-row photon count discriminant. 
This suggests that incorporating the full spatial information provides additional discrimination power beyond considering only simple variables describing the $z$ dependence of the light distribution, though the $z$-dependent component captures the majority of the achievable rejection.

\subsection{Scintillation Based Bi-Po Tagging}
In LXe, $\alpha$ particles have a short range and deposit energy densely, producing highly localized scintillation signals. 
The detectability of these $\alpha$ signals can be estimated from the expected detected photon yield,
$N_{\rm det} = N_{\rm scint} \times {\rm LCE}.$
For a \qty{7.7}{MeV} $\alpha$ depositing its full energy in LXe, the light yield determined from NEST gives an average of approximately $9\times10^{5}$ scintillation photons. 
With the estimated below-cathode LCE of 3.2\%, this corresponds to $\sim 3\times10^{4}$ detected photons. 
As a conservative benchmark, we consider a simulated event in which the $\alpha$ is emitted from an implanted nucleus and deposits only \qty{2}{\MeV} in the LXe (Fig.~\ref{fig:alpha_light}).
NEST predicts approximately $2.4\times10^{5}$ scintillation photons for this case, corresponding to $\sim 8\times10^{3}$ detected photons.
These signals are well above the photon counts expected from SiPM dark noise and therefore are readily taggable.

MC simulations indicate that half of \po{} $\alpha$ decays deposit near their full energy in the LXe rather than the cathode material. 
Combined with the estimates of the LCE below the cathode, this implies a bottom surface Bi-Po tagging efficiency of $\sim$50\%, substantially larger than the conservative $<$1\% assumed in previous nEXO sensitivity studies~\cite{Adhikari_2021}.
This brings the expected tagging performance for bottom-surface decays into the same regime as that of top-surface decays.

\section{Discussion}
While the simulations presented here incorporate a number of improvements over previous work, several limitations of the present study should be noted.
Owing to the limited sample of signal-like background events, a relatively simple network architecture was adopted, and additional discrimination power may be achievable with larger simulated datasets and more sophisticated machine-learning approaches.
We have assumed optimistic noise levels for the charge readout and increased noise would reduce the efficiency of reconstructing low-energy $\beta$ deposits near the cathode, though the more effective prompt scintillation-based tagging and below-cathode Bi-Po tagging results are largely independent of charge reconstruction performance.
Finally, uncertainties in the fluid flow and surface attachment of radon daughters may affect the assumed fraction of background events on the cathode, though the conceptual tagging mechanisms described in this work are still applicable.

Assuming that the prompt and delayed scintillation-based tagging methods act independently, the rejection efficiencies reported in Section~\ref{sec:taggingperformance} at 95\% signal efficiency can be naively combined. Applying these to the cathode contribution of approximately 33\% of the total background budget \cite{Adhikari_2021}, with the top and bottom surfaces weighted by their contributions after the existing Bi-Po veto, gives an estimated reduction in the total background of approximately 16\% using the bottom-row photon count and Bi-Po tagging, increasing to approximately 21\% with the CNN classifier.
These rejection efficiencies are evaluated as cuts on individual observables and are not incorporated into the full multiparameter likelihood framework used in the nEXO sensitivity analysis. Consequently, the impact on the final \nldbd\ sensitivity remains to be quantified in detail. 

Radon progeny backgrounds are expected to remain an important consideration for future tonne-scale liquid xenon detectors. Though we have focused on the application of these tagging techniques to the specific single-phase nEXO TPC design, the rejection techniques explored here rely on general features of the Bi-Po decay sequence and the spatial distribution of associated scintillation signals, and could therefore also be implemented in other LXe TPC detectors. 
Two-phase LXe detectors can achieve substantially lower charge-detection thresholds through electroluminescence amplification in the gas phase, reaching sensitivities to single ionization electrons \cite{Edwards:2008singleelectron, XENON:2025ionization, LZ:2026fewelectron}. The ability to detect small amounts of charge emitted from the top of the cathode implies that the charge-based $\beta$ tagging methods described in Section~\ref{sec:chargebeta} are likely to be much more effective than in a single-phase design. 
Additionally, in both single and two-phase detector designs, the light-based tagging techniques developed here offer complementary discriminating power for events occurring in regions where charge is not fully collected. It should be noted that the specific light-tagging algorithms used in this work were optimized for the highly pixelated photon hit patterns afforded by nEXO's optically open field cage and barrel mounted SiPM array, which extends below the cathode. These algorithms would need to be similarly tailored to LXe TPC designs that employ photosensor arrays mounted only at the two ends of the drift volume with an optically open cathode grid. 
In addition to the rejection techniques explored here, the recoil implantation study provides an estimate of the depth distribution of radon progeny within surfaces. Accurate implantation profiles are needed for predicting the escape and detection probabilities of $\alpha$ particles, $\beta$ particles, and nuclear recoils from radon progeny decays, and may therefore be useful for future studies of radon-induced backgrounds in noble-liquid detectors \cite {XLZD:2024pdv}.

\section{Conclusions}
In this work we have presented novel ideas to tag \bi\ decays from radon progeny implanted within the detector surfaces of LXe TPCs through the detection of correlated energy deposits. 
By incorporating a physically motivated model of radon progeny transport and implantation, we have shown that charge and light signals from the coincident $\beta$ decay or the subsequent $\alpha$ decay can be used to discriminate \bi\ \qty{2448}{keV} $\gamma$-rays from a potential \nldbd\ signal.

Charge reconstruction of the coincident $\beta$ particle provides discrimination for top-surface decays, but is much less effective for bottom-surface decays where ionization charge is not collected. 
After the application of all standard nEXO reconstruction selections and the inner 2T fiducial volume requirement, a residual population of signal-like background events remains for which scintillation-based observables provide additional discrimination. A simple bottom-row photon count rejects 41\% of the remaining \bi{} backgrounds at 95\% signal efficiency, while a CNN-based classifier improves this rejection to 57\%.

The updated estimate of the light collection efficiency below the cathode increases the expected Bi-Po tagging efficiency for bottom-surface decays from the previously assumed value of less than 1\% to approximately 50\%, now comparable to the tagging efficiencies of top-surface decays.
Together, these results indicate that a higher fraction of cathode-implanted \bi{} backgrounds can be rejected than previously assumed and motivate the incorporation of scintillation-based observables, already available within the nEXO detector design, into future sensitivity projections.

\section{Acknowledgements}
M.P.W. is supported by the National Science Foundation Graduate Research Fellowship under Grant No. DGE-2139841. B.T.F. and R.S. were supported by the U.S. Department of Energy Office of Science, Office of Nuclear Physics Early Career Research Program. The authors gratefully acknowledge support for nEXO from the Office of Nuclear Physics within DOE’s Office of Science under Grants/Contracts No. DE-AC02-76SF00515, No. DE-FG02-01ER41166, No. DE-SC002305, No. DE-FG02-93ER40789, No. DE-SC0021388, No. DE-SC0012704, No. DE-AC52-07NA27344, No. DE-SC0017970, No. DE-AC05-76RL01830, No. DE-SC0012654, No. DE-SC0021383, No. DE-SC0014517, No. DE-SC0024666, No. DE-SC0020509, and No. DE-SC0024677, and by the U.S. National Science Foundation under Grants No. NSF PHY-2111213 and No. NSF PHY-2011948; from NSERC SAPPJ-2022-00021, CFI 39881, FRQNT 2019-NC-255821, and the CFREF Arthur B. McDonald Canadian Astroparticle Physics Research Institute in Canada; from IBS-R016-D1 in South Korea; and from CAS in China. This work was supported in part by Laboratory Directed Research and Development (LDRD) programs at Brookhaven National Laboratory (BNL), Lawrence Livermore National Laboratory (LLNL), Pacific Northwest National Laboratory (PNNL), and SLAC National Accelerator Laboratory.

\bibliographystyle{unsrt_et_al}
\bibliography{references_cleaned_nima}

\end{document}